\documentclass[useAMS,usenatbib]{mn2e}
\usepackage{color}
\usepackage{graphicx}
\usepackage{srcltx}
\title[Automated classification pipeline for TAUVEX]
{A 3D Automated Classification Scheme for the TAUVEX data pipeline}
\author[Bora et al.]{Archana Bora$^{1,2}$\thanks{E-mail:
archana@iucaa.ernet.in; rag@iucaa.ernet.in; hpsingh@physics.du.ac.in;
jmurthy@yahoo.com reks@iiap.res.in and kalpanaduorah@yahoo.com },
Ranjan Gupta$^{1}$, Harinder P. Singh$^{3}$, Jayant Murthy$^{4}$
\newauthor Rekhesh Mohan$^{4}$ and Kalpana Duorah$^{2}$ \\
$^{1}$IUCAA, Post Bag 4, Ganeshkhind, Pune-411007, India\\
$^{2}$Department of Physics, Gauhati University, Guwahati 781014, India\\
$^{3}$Department of Physics \& Astrophysics, University of Delhi,
         Delhi 110007, India\\
$^{4}$Indian Institute of Astrophysics, Koramangla, Bangalore 560034, India\\
}
\begin{document}

\date{Received on 15/6/2007; Accepted on 23/11/2007}

\pagerange{\pageref{firstpage}--\pageref{lastpage}} \pubyear{2007}

\maketitle

\label{firstpage}

\begin{abstract}
In order to develop a pipeline for automated classification of stars to be observed by the TAUVEX ultraviolet space Telescope, we employ an artificial neural network (ANN) technique for classifying stars by using synthetic spectra in the UV region from 1250\AA~ to 3220\AA~ as the training set and International Ultraviolet Explorer (IUE) low resolution spectra as the test set. Both the data sets have been pre-processed to mimic the observations of the TAUVEX ultraviolet imager. We have successfully classified 229 stars from the IUE low resolution catalog to within 3-4 spectral sub-class using two different simulated training spectra, the TAUVEX spectra of 286 spectral types and UVBLUE spectra of 277 spectral types. Further, we
have also been able to obtain the colour 
excess (i.e. E(B-V) in magnitude units)
or the interstellar reddening for those IUE spectra which 
have known reddening to an accuracy of better than 0.1 magnitudes. It has been shown that even with the limitation of data from just photometric bands, ANNs have not only classified the stars, but also provided satisfactory estimates for interstellar extinction. The ANN based classification scheme has been successfully tested on the simulated TAUVEX data pipeline. It is expected that the same technique can be employed for data  validation in the ultraviolet from the virtual observatories. Finally, the interstellar  extinction estimated by applying the ANNs on the TAUVEX data base would provide an extensive extinction map for our galaxy and which could in turn be modeled for the dust distribution in the galaxy.

\end{abstract}


\section{Introduction}

Tel-Aviv University Ultra-Violet Experiment (TAUVEX) is an Indo-Israeli Ultraviolet Imaging space mission that will image large parts of the sky in the wavelength region between 1300 and 3200\AA. The instrument consists of three equivalent 20-cm UV imaging telescopes with a choice of filters for each telescope. Each telescope has a field of view of about 54' and a spatial resolution of about 6" to 10", depending on the wavelength. TAUVEX will be launched into a geostationary orbit as part of Indian Space Research Organization's GSAT-4 mission in April 2008.

Observations will be available using filters in five UV bands:
\begin{enumerate}
\item BBF     : Broadband filter (1300--3300\AA)
\item SF1     : Intermediate band filter 1 (1250--2250\AA)
\item SF2     : Intermediate band filter 2 (1800--2600\AA)
\item SF3     : Intermediate band filter 3 (2100--3100\AA)
\item NBF     : Narrowband filter (2000--2400\AA)
\end{enumerate}

Figure 1 shows the response curves for each of the TAUVEX filters
in units of Effective Area $cm^{2}$.

\begin{figure}
\centering
\includegraphics[width=8.5cm]{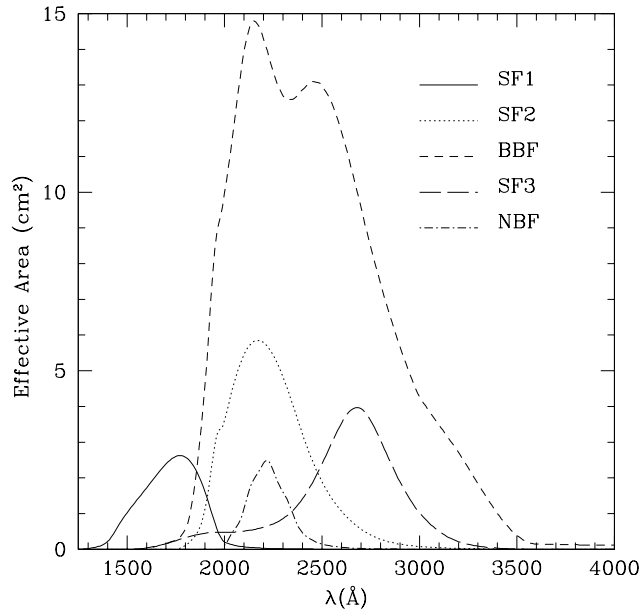}
\caption{Filter response of the five filters of TAUVEX}
\label{Fig1}
\end{figure}

The TAUVEX mission will
have added advantages as compared to other earlier UV missions like the
TD satellites and GALEX etc. 
The estimation of the slope $\rm R_{v}$ of the interstellar 
extinction curve with a greater sensitivity, will allow to construct deeper
maps of the UV sky.
Further, TAUVEX and TD satellites would complement each other by having
a total of six data points for the interstellar 
extinction curve for their common sources (see Maheshwar et al. 2007).

TAUVEX will mostly operate in scanning mode, since it will be mounted on GSAT-4, a geosynchronous satellite. The FOV will be scanning a strip of the sky with constant declination and 
a limiting magnitude of 19 (Murthy, 2003). A few years of successful run of the mission will record more than a million UV point sources apart from galaxies, QSO's and the UV background. The need for an automated classification pipeline for the stellar sources that is
repeatable and fast is, therefore, immense.

The Artificial Neural Network (ANN) based schemes are now being routinely used to classify spectra from large spectral data bases (Gulati et al. 1994, Singh et al. 1998, 2006, Valdes et al. 2004, Bailer-Jones 2002, Gupta et al. 2004) for the purpose of sorting these large spectral data base into groups of main spectral types (O, B, A, F, G, K and M) and sub-classes. Further, these schemes can also be used for obtaining stellar fundamental atmospheric parameters (Gulati et al. 1997 a,b). Of these Gulati et al. (1997b) is of particular interest, since it was shown that ANNs can determine the colour excess, i.e. E(B-V) in units of magnitudes, as an additional parameter when applied to the IUE spectral data base.

The current work has used the ANN based tools for classifying the IUE 
spectral data base (reduced to the TAUVEX band data) in terms of the spectral 
types and also hierarchically estimated the color excess using
this tool. It is worth noting that whereas 
the Gulati et al. (1997b) used the IUE full spectra for spectral type classification 
and estimation of colour excess, the present work uses the simulated band data as 
expected from the TAUVEX satellite and even with this limitation, the neural 
network scheme has been able to assign the spectral classes and also obtain 
reddening estimate to satisfactory levels.

In the next section, we describe the generation and pre-processing of the simulated TAUVEX data that is used for training of the neural network as well as the processing of the IUE spectral data which is used as the test set. In Section 3, we describe results of the ANN classification scheme as well as the color excess determination. In Section 4, we present important conclusions of the study.

\section[]{ANN architecture, generation of simulated data and ANN train and test sets}

Following sub-sections describe the ANN architecture, simulated data generation and the ANN train and test sets.

\subsection[]{ANN architecture}
The ANN architecture considered here is an supervised one with a minimum configuration of three layers, i.e., (1) Input layer where the patterns are read (2) Hidden layer where the information is processed from the input layer (3) Output layer where the output patterns are rendered (see Bailer-Jones et al. 2002 for a review). The hidden layer can have several nodes which inter-connect the input and output layers with each connection with its designated
connection weight. We have used a back-propagation algorithm (Gulati et al. 1994, 1997a,b, Singh et al. 1998) with 2 hidden layers of 64 nodes each and this scheme requires a training session where the ANN output and the desired output get compared after each iteration and
the connection weights get updated till the desired minimum error threshold is reached. At this stage the network training is complete and the connection weights are considered frozen. The next stage is the testing session where the test patterns are fed to the network and output is the classified spectral pattern or color excess
in terms of the training sets.

In the actual post launch of TAUVEX when the real data 
will be available, the 
scheme applied to estimate the colour excess will have to run the ANN in two 
stages i.e. in a hierarchal manner such that, the first stage classifies the 
test set (IUE data base or the expected TAUVEX data base) into the spectral classes 
and then a second ANN stage performs the colour excess estimation.

\subsection[]{Simulated data generation}

We have used two independent sources to generate the training sets of spectra 
with solar type stars with $[M/H] = 0$. One is the stellar flux calculator
from TAUVEX website (http://tauvex.iiap.res.in/htmls/tools/fluxcalc/) containing 286 spectro-luminosity classes and the other is the UVBLUE fluxes (Rodriguez-Merino et al. 2005)
(http://www.bo.astro.it/$\sim$eps/uvblue/uvblue.html). 
Based on the spectral type and luminosity class of a star, the TAUVEX calculator
derives the effective temperature and surface gravity using the calibration of
Allen (2000), Colina (1995) and Lang (1982) and 
calculates the spectral energy distribution for each star using appropriate 
Kurucz model available on the webpage http://kurucz.harvard.edu/ (see
Sujatha et al. 2004).
We have used the information from Allen (2000), Erika B\"e{o}hm-Vitense (1981), 
Johnson (1966), Ridgway et al. (1980), Alonso et al. (1999) and
Bertone et al. (2004) for matching the parameter space of UVBLUE to spectral-types 
and luminosity classes. 
Both the sources provide sets of theoretical 
fluxes (based on Kurucz model atmospheres) in the UV region. 
These fluxes need to be processed via a common flux integration programme
provided at the TAUVEX tools site to form two sets of band data 
(each having four fluxes corresponding to the four TAUVEX bands) 
and they constitute the simulated
 band data set for the ANN training sets.

We have also obtained two sets of fluxes (with 50\AA~ resolution and 40 data bins 
covering the spectral region of 1250--3220\AA) aimed at preparing the ANN 
tools for another Indian scientific mission satellite ASTROSAT 
(http://www.rri.res.in/astrosat/) which will have gratings to provide 
slit-less spectra for spatially resolved stars. It will also prepare us for 
the future GAIA mission 
(http://gaia.esa.int/science-e/www/area/index.cfm?fareaid=26).

\subsection[]{ANN train and test sets}

While making the train and test sets, one has to ensure that the number of spectral 
fluxes at the respective wavelengths and the starting/ending wavelengths are 
identical. Also the spectral resolution needs to be same and for this, the 
spectral libraries had to be convolved with appropriate gaussian functions to 
bring them at par with each other. The fluxes are normalized to 
unity with respect to maximum flux in each spectrum before sending to the ANN
inputs.  The spectra for 286 TAUVEX spectral types generated in the
range 1250-3200\AA~ have a resolution of 10\AA~ which we have degraded to
50\AA~. The resolution of 277 UVBLUE spectral types have been degraded similarly
(using the relevant codes provided on the UVBLUE library web site). 
These sets of data are then reddened (using the observed extinction curve of
Seaton, 1979) in the range of 0.00-1.00 mag, for preparing the training sets for the 
two stages of the hierarchal scheme viz. the separation of the different spectral types and 
the evaluation of reddening values. Below we provide
the details of the procedure adopted for generating the training sets for the 
two stages:

\begin{itemize}

\item {\bf Generating data set for Spectral Type determination:}

In the first stage, reddening values are added in step sizes
of 0.20 magnitudes to the simulated data. The 0.20 step is chosen for the computational
convenience. For example, the TAUVEX data consists of 286 different classes with 58
spectral types, each having 5 luminosity classes (except for O6.5V).
If one wants to classify the Spectral type, Luminosity  
class and the reddening value in a single run;
reddening these 286 data sets with reddening value from 0.00-1.00,
even at a step of 0.1 leads to $286 \times 11$= 3146 number of distinct classes.
However, this is not possible 
with our current computational facilities and the present version of our ANN.
Instead, we go for the hierarchal scheme by first merging all the luminosity 
classes.
For example, instead of considering O3I-O3V as five separate classes, the ANN will
be trained to learn all the five different patterns as single O3 spectral type only,
though the variation in all the five spectra still go as input to the ANN. 
The process thus reduces the number of distinct classes from 286 to 
only 58 classes, making the computation fast.
When the learning process is completed, 
ANN can separate different Spectral types, thus
making it possible to find out the reddening values in the next stage.
 
\item {\bf Generating data set for Reddening} evaluation:

In the second stage, reddening values are added in step
sizes of 0.05 to the simulated data. The separation of the available spectra
into different groups O, B, A, F, G, K etc. in the first stage, makes it possible
to select this finer step size of 0.05. In our work we have not classified the 
luminosity classes separately, however, this can be done easily by adding
one more stage in the hierarchal scheme.
\end{itemize}

A sample of normalized simulated spectra of different spectral types is shown in 
Fig. 2. Their integrated fluxes in the four TAUVEX bands, 
NBF, SF1, SF2 and SF3, have been computed using respective filter response curves of
Fig.1. Fig. 3 shows the residuals obtained by subtracting IUE fluxes from the
corresponding TAUVEX simulated fluxes. The discrepancies observed in these figures
could be due the following reasons:

In the early-type stars i.e. O and B, the main discrepancy
between observed and theoretical is near 1500\AA. 
This is a consequence of the physical origin of the C{\small {IV}} line, 
which gets strongly 
affected by stellar winds and mass-loss processes in massive stars.
For F-type stars the metallic features at 2400\AA(Fe III), 2500\AA(Fe I/Si I),
2800\AA(Mg II) are more enhanced in the simulated spectra.
For G-type stars the chromospheric activities increases and thus triggers
prominent Mg core emissions which are not seen in the simulated spectra. 
The chromospheric activities are not accounted for in the Kurucz's model
(Rodriguez-Merino et al. 2005). The discrepancy is more clearly visible in 
the band integration of the fluxes of late type stars in Fig. 3.

The final training set thus contains (a) the spectra in the form shown in 
Fig. 2 and (b) 4 flux values in the 4 bands of TAUVEX in the form shown in 
Fig. 3 -- for each of the 286 TAUVEX spectra (277 spectra for the UVBLUE case) 
with reddening in the range of 0.00 to 1.00 mag with a step of
0.2 mag. Fig. 4 shows a block diagram of the flow chart for 
preparing these two training sets for spectral type 
classification.

\begin{figure}
\centering
\includegraphics[width=8.5cm]{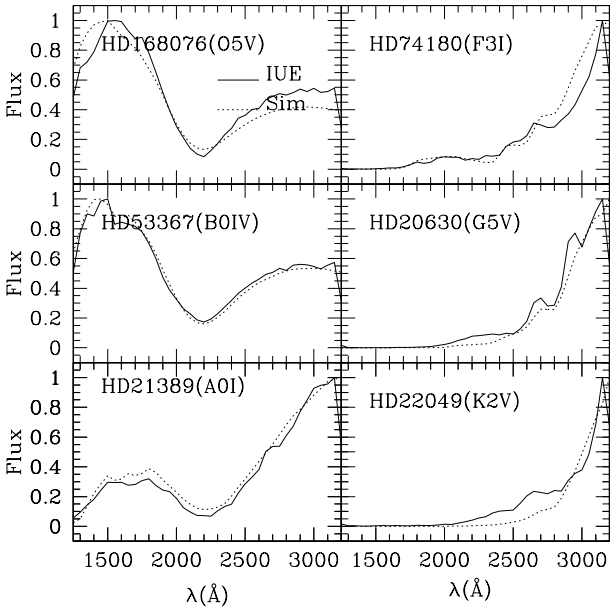}
\caption{IUE and TAUVEX Simulated fluxes for 6 sample stars at a resolution
of 50\AA.}
\label{Fig2}
\end{figure}

\begin{figure}
\centering
\includegraphics[width=8.5cm]{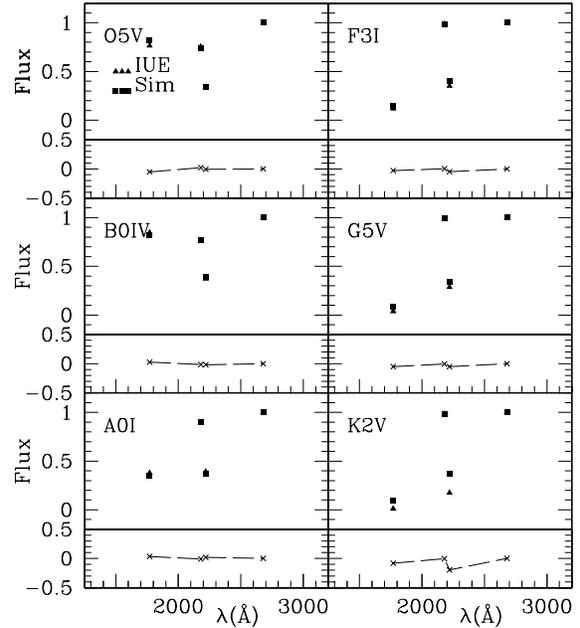}
\caption{Integrated IUE and Simulated TAUVEX fluxes for the same 6 sample stars in 
NBF, SF1, SF2 and SF3 filters along with the residues in the corresponding lower panels.}
\label{Fig3}
\end{figure}

\begin{table*}
\begin{center}
\caption{Number of Spectra for each data set according to the spectral types.}
\begin{tabular}{lccc}
\hline
Spectral Class & TAUVEX & UVBLUE & IUE \\
\hline
O  & 36 & 36 & 42 \\
B & 50 & 41 & 115 \\
A & 50 & 50 & 48 \\
F & 50 & 50 & 20\\
G & 50 & 50 & 3 \\
K & 50 & 50 & 1\\
\hline
\end{tabular}
\end{center}
\end{table*}

\begin{figure}
\centering
\includegraphics[width=8.5cm]{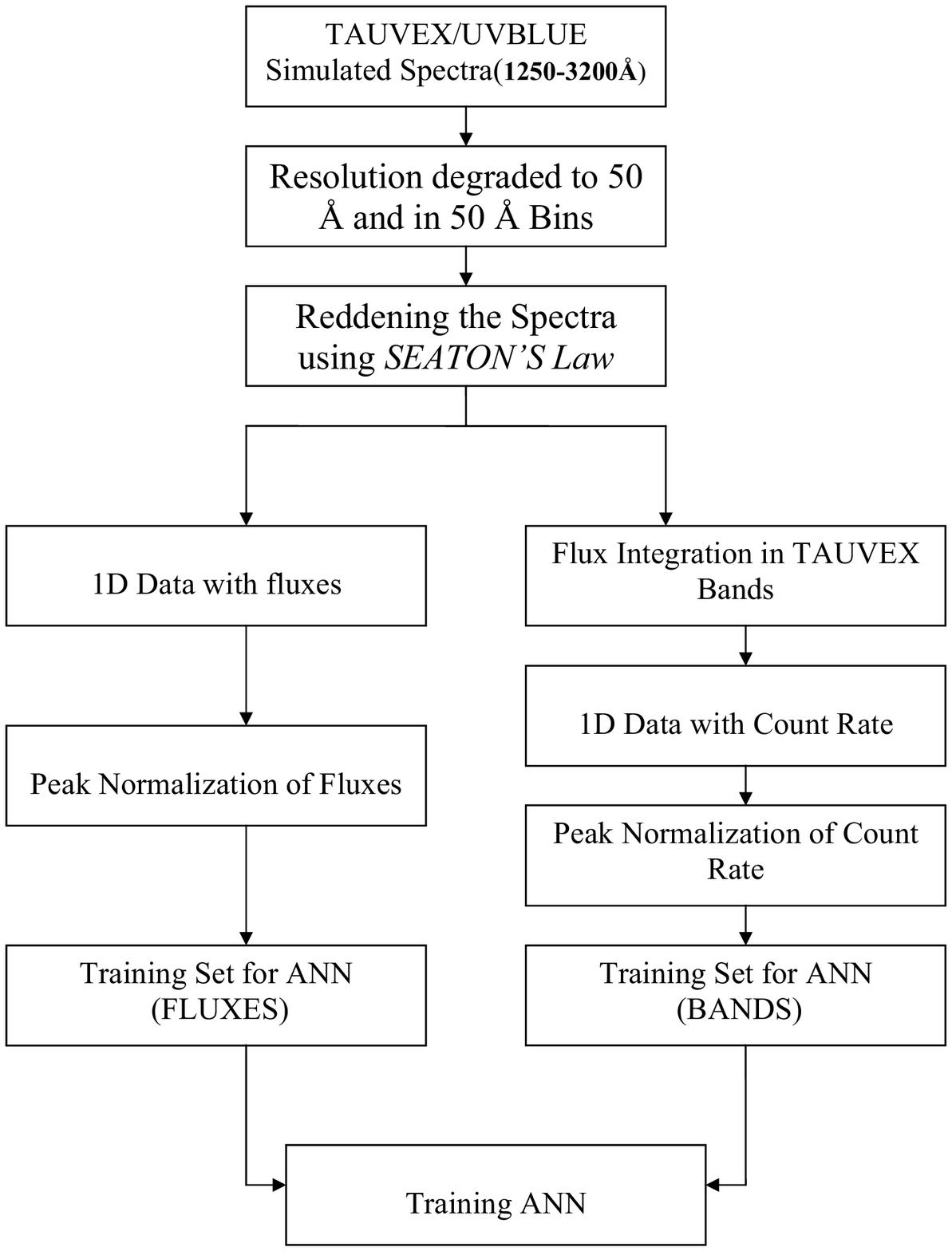}
\caption{A block diagram showing the flow chart for creating the ANN
train set for spectral classification with TAUVEX
and UVBLUE simulated sources.}
\label{Fig4}
\end{figure}
The test spectra  were taken from the IUE low resolution spectra:
reference atlas, normal stars, ESA SP-1052 by Heck et al. (1984)
which contains 229 low-dispersion flux calibrated spectra of O to K
spectral type obtained by the IUE satellite. The spectra were
trimmed to 1250--3220\AA~. The original resolution of 6\AA~ of IUE
spectra was convolved by a Gaussian function to produce a degraded
resolution of 50\AA~. Fig. 5 shows the block diagram of the flow
chart for generating this IUE test set for spectral classification. 
The Fig. 6 shows a block diagram for the flow chart for creating the train set
for extinction classification and Fig. 7 shows the corresponding block diagram 
of the flow chart for creating the IUE test set.

\begin{figure}
\centering
\includegraphics[width=8.5cm]{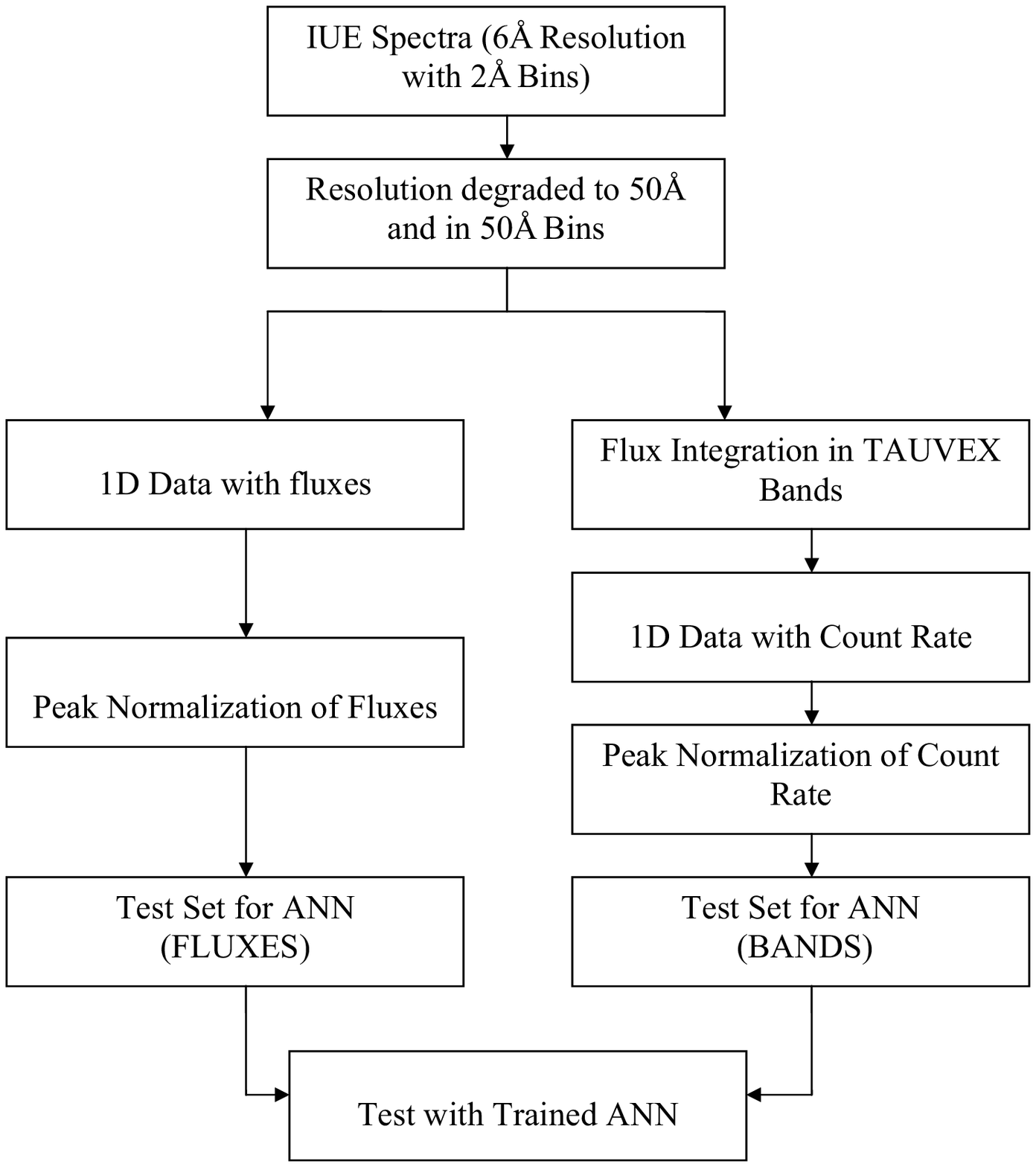}
\caption{A block diagram showing the flow chart for creating the ANN
test set for corresponding to the Figure 4.}
\label{Fig5}
\end{figure}

Table 1 shows the number of spectra per spectral type used in this
analysis. 
The numbers in the 2nd and 3rd column are the 
basic sets for traning 
sessions of the ANN. The hierarchal ANN scheme used by 
us works in two stages viz. Ist stage performs the spectral type classification
and for this these numbers get multiplied by 6 and in the IInd stage which
performs the color excess classification, they get multiplied by 21.
Further, in order to have an unifrom number of spectra per spectral type,
classes which have just one example are duplicated during the training session.

\begin{figure}
\centering
\includegraphics[width=8.5cm]{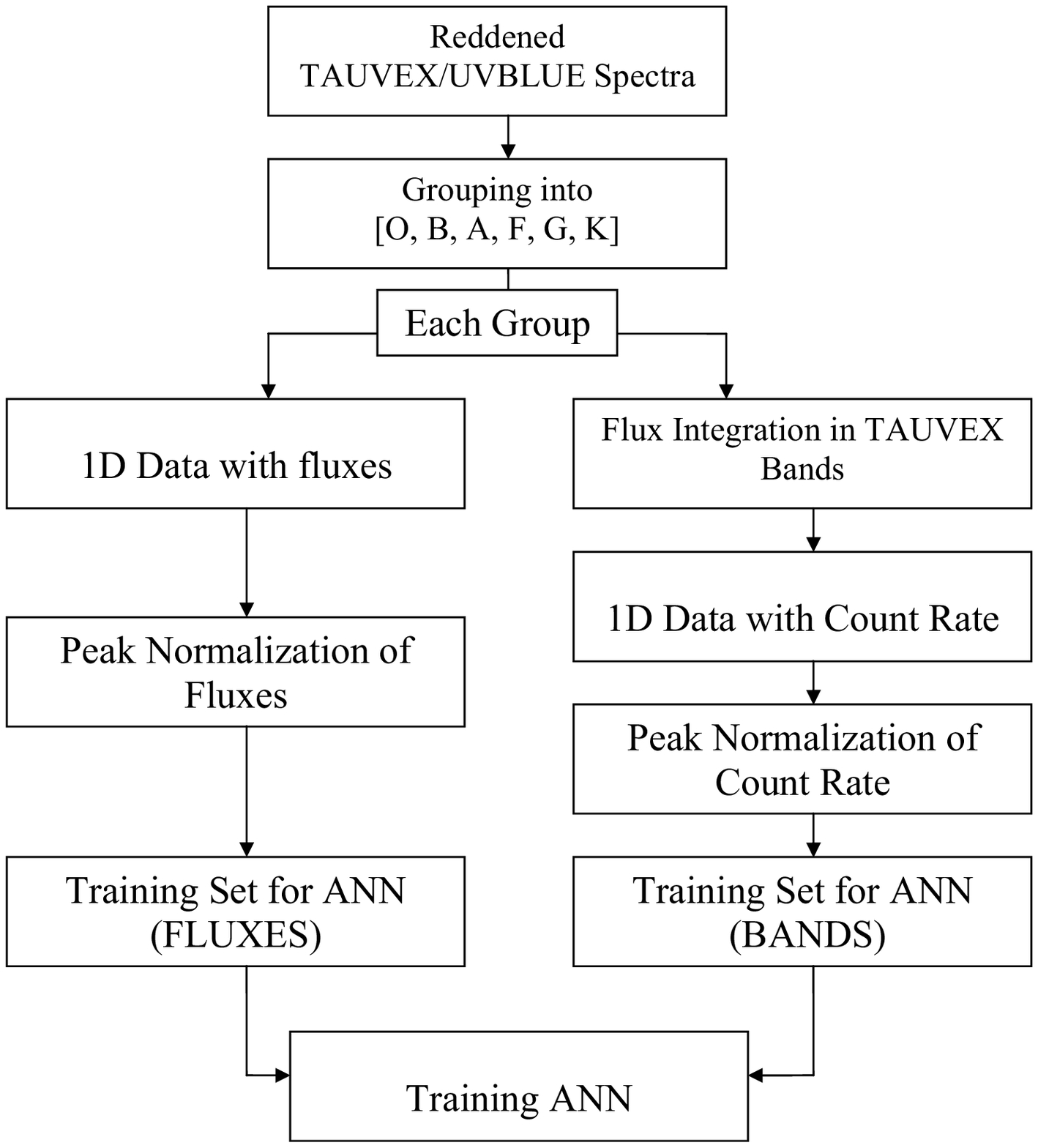}
\caption{A block diagram showing the flow chart for creating the ANN
train set for extinction classification for both simulated sources
i.e. TAUVEX and UVBLUE.}
\label{Fig6}
\end{figure}

\begin{figure}
\centering
\includegraphics[width=8.5cm]{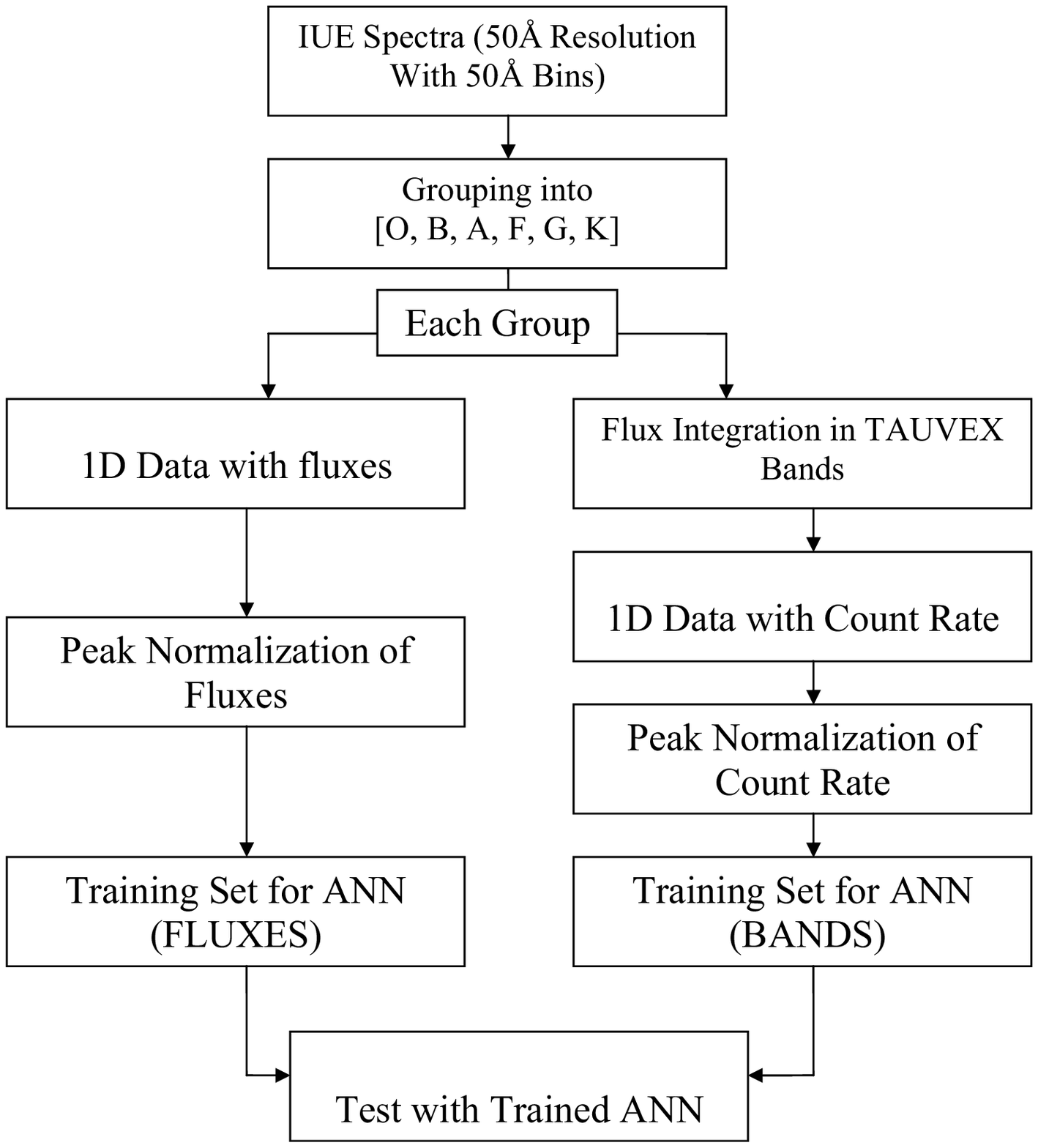}
\caption{A block diagram showing the flow chart for creating the ANN
test set for extinction classification corresponding to the Fig. 6.}
\label{Fig7}
\end{figure}

\section[]{Results of the ANN Classification}

The results of spectral classification are depicted in the Fig. 8. The numbers on 
the axes of this figure refer to the spectral coding which is briefly described 
as follows:\\

\noindent Main Spectral Type: O = 1000, B = 2000, A = 3000,
.......... K =
6000,\\

\noindent Sub-Spectral Type: O1 = 1100, O2 = 1200, ........... O9 =
1900,\\

\noindent Luminosity Class: I = 1.5, II = 3.5, III = 5.5, IV = 7.5
and V =
9.5.\\

\noindent For example, Sun is a G2V star and  hence its code will be
5209.5. A Classification error of 500 implies that a G2 star can, at
worse, be classified either as F7 or G7 spectral type.

\begin{figure}
\centering
\includegraphics[width=8.5cm]{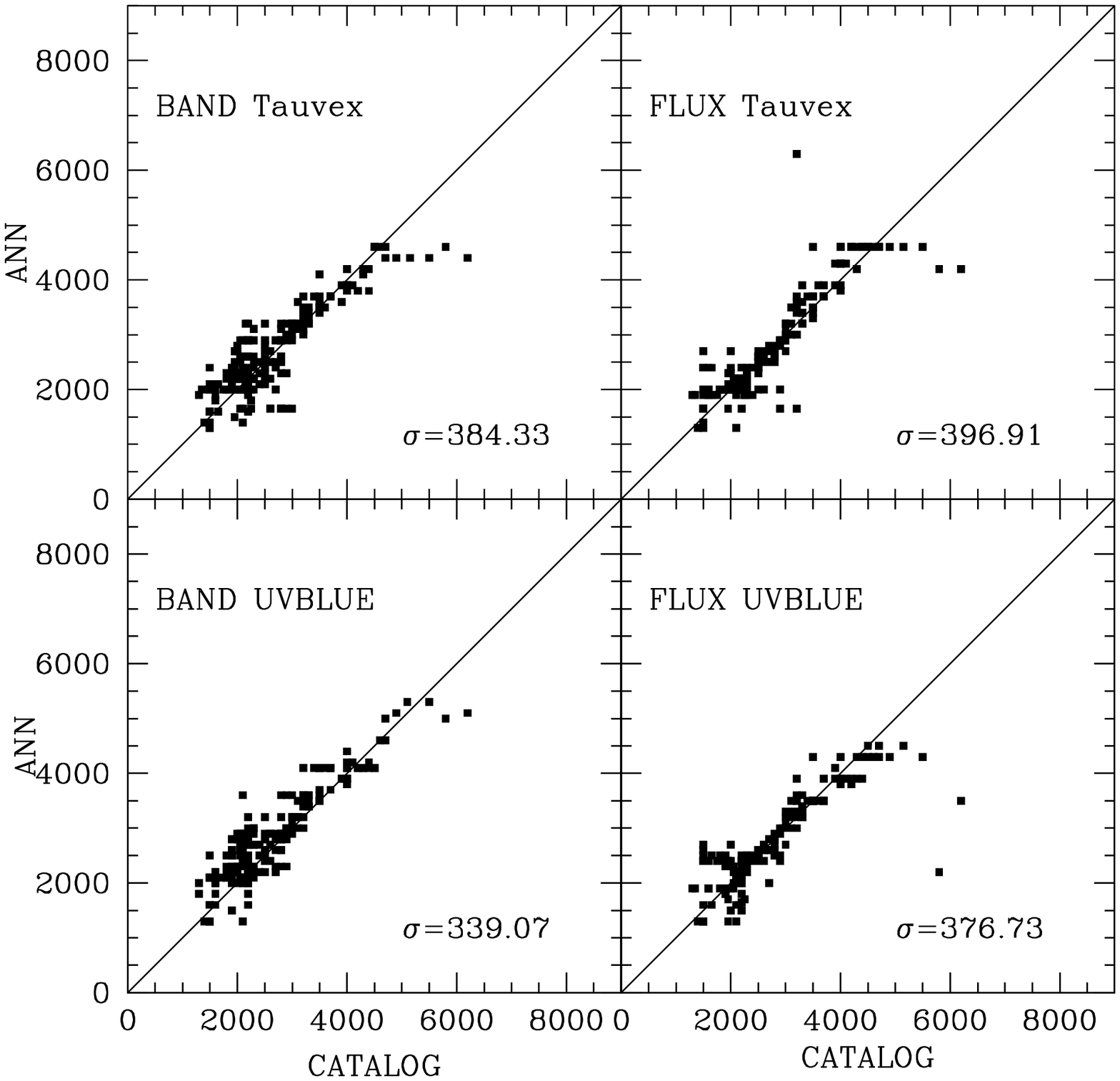}
\caption{Scatter plots of classification of the 229 IUE stars with TAUVEX
bands and fluxes and with UVBLUE bands and fluxes. The classification
accuracy values $\sigma$ are shown for each case  
in units of sub-spectral types.}
\label{Fig8}
\end{figure}

Figure 9 shows the scatter plots for pre-classified IUE stars (in O, B, A and F spectral types) for UVBLUE fluxes with their colour excess estimates $\sigma$ in units of magnitudes. Figure 10 shows the scatter plots for pre-classified IUE stars (in O, B, A and F spectral types) for UVBLUE bands with their colour excess estimates $\sigma$ in units of magnitudes. Figures 11 \& 12 show the corresponding classification results for TAUVEX fluxes and bands respectively. 
In these 3D scatter plots, the 'Cat' and 'ANN' denote the
catalog and ANN classes respectively. Further, the vertical axis in the plots 
gives the number of stars (N) present for a particular color excess value and 
are re-scaled as the square root of the actual number (i.e. $\rm N^{1/2}$) for 
better representation; otherwise in the cases where this 
number is large, the corresponding points for single stars would look too small 
on the plots. 

It is important to see that in the spectral classification
scheme, the outliers in the all the four panels of Fig. 8 belong to G and K type, 
they being misclassified as the F type stars. This can be attributed to the
discrepancies mentioned in section 2.3. In the two exceptional cases 
G8 gets classified as O2 type in FLUX UVBLUE panel whereas A2 gets classified 
as K3 in FLUX TAUVEX panel. The misclassification of G8 as O2 may be because as G8 
IUE spectra shows a moderate UV excess compared to the theoretical one as 
mentioned in Rodriguez-Merino et al. (2005).

\newcommand{\boxfig}[2]
{\fbox{\begin{minipage}{1.5in}
\centering
\includegraphics[totalheight=1.5in, width=1.5in]{#1.eps} \\
#2
\end{minipage}}}

\begin{figure}
\centering
\boxfig{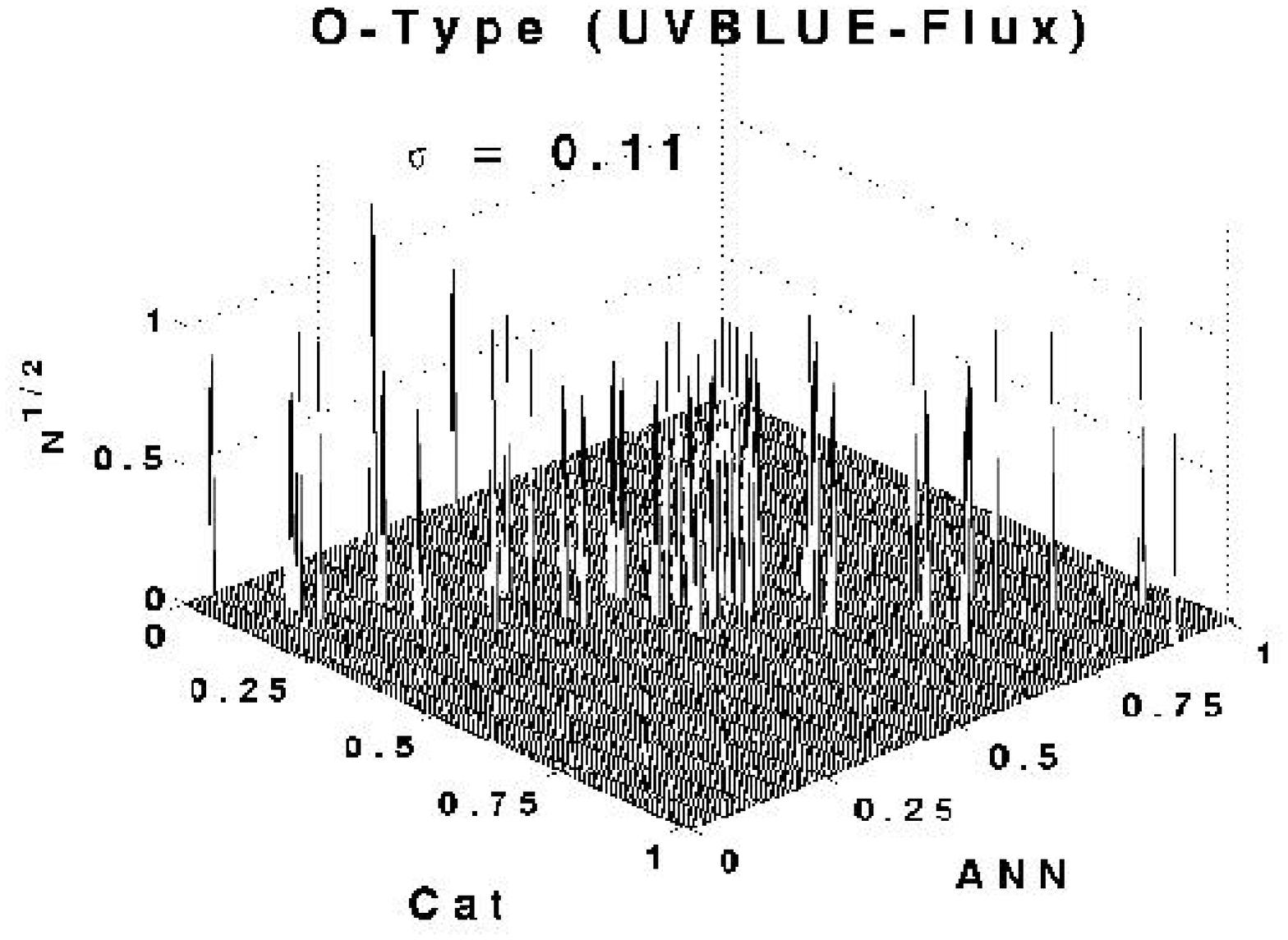}{}
\boxfig{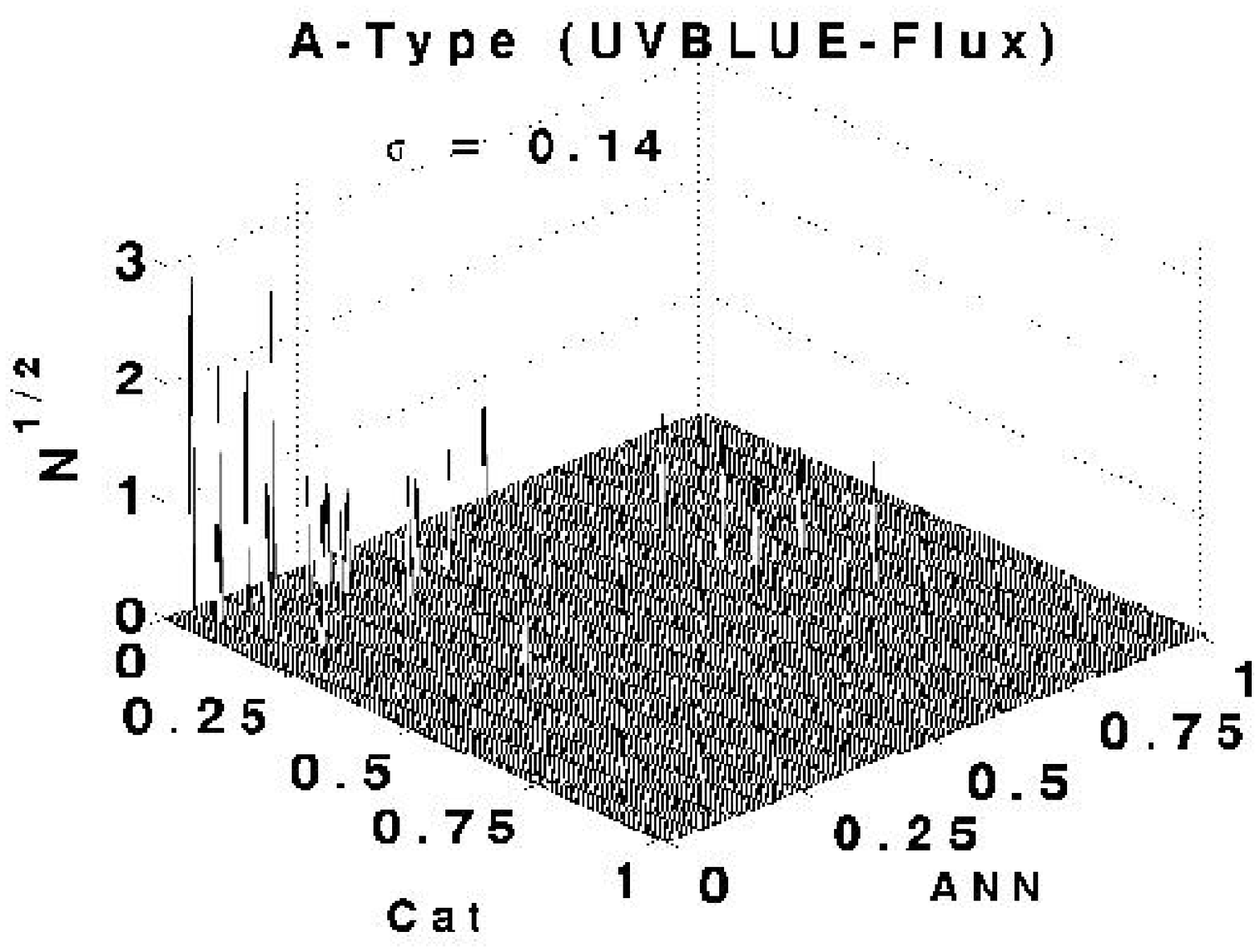}{}
\hspace{0.05in}
\boxfig{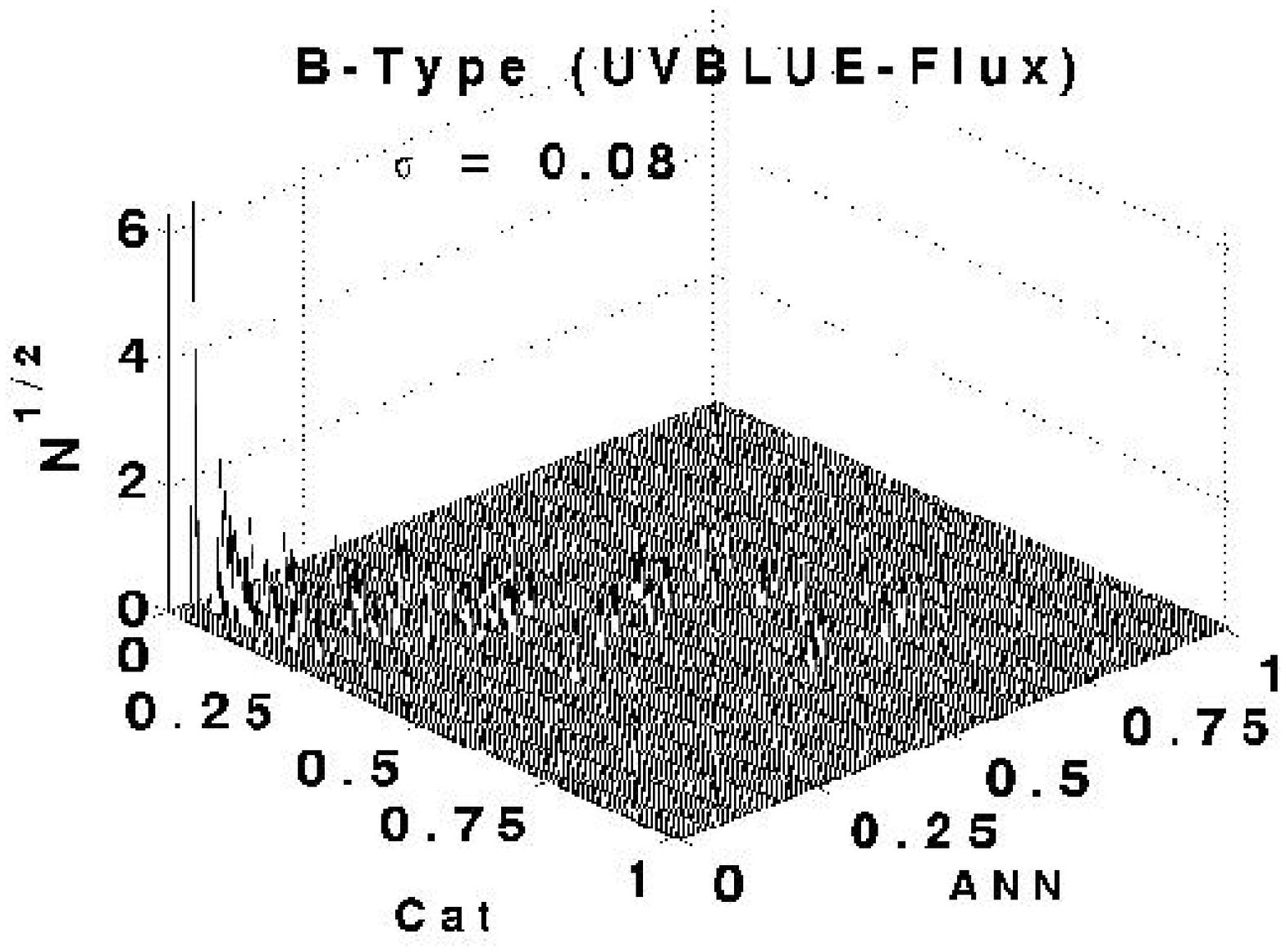}{}
\boxfig{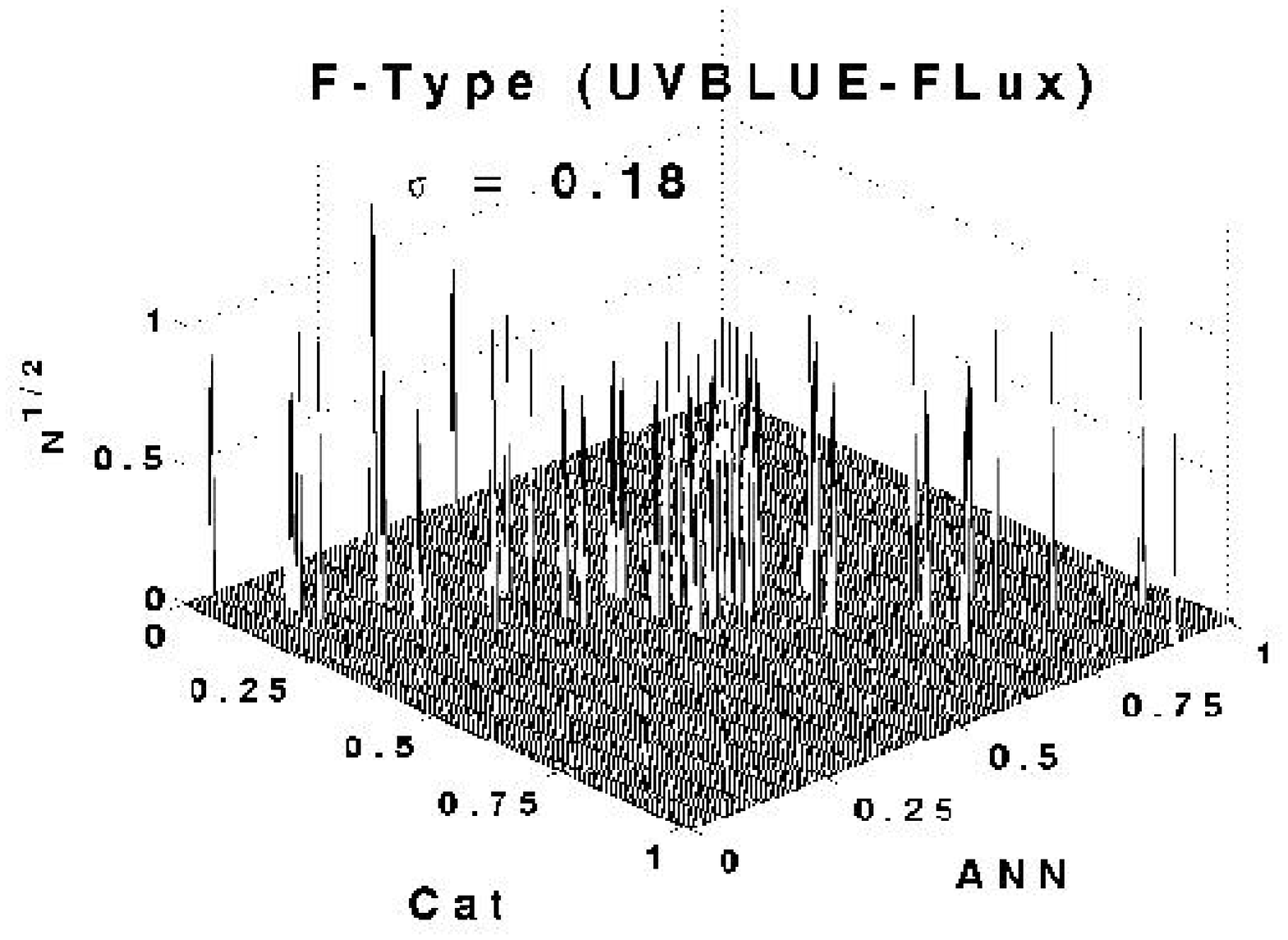}{}
\caption{Scatter plot of classification of 229 IUE stars (pre-classified
 into O, B, A and F spectral types) with UVBLUE fluxes for colour excess estimates. The
 classification accuracy values $\sigma$ are shown for each case in units of
E(B-V) magnitudes.}
\end{figure}

\begin{figure}
\centering
\boxfig{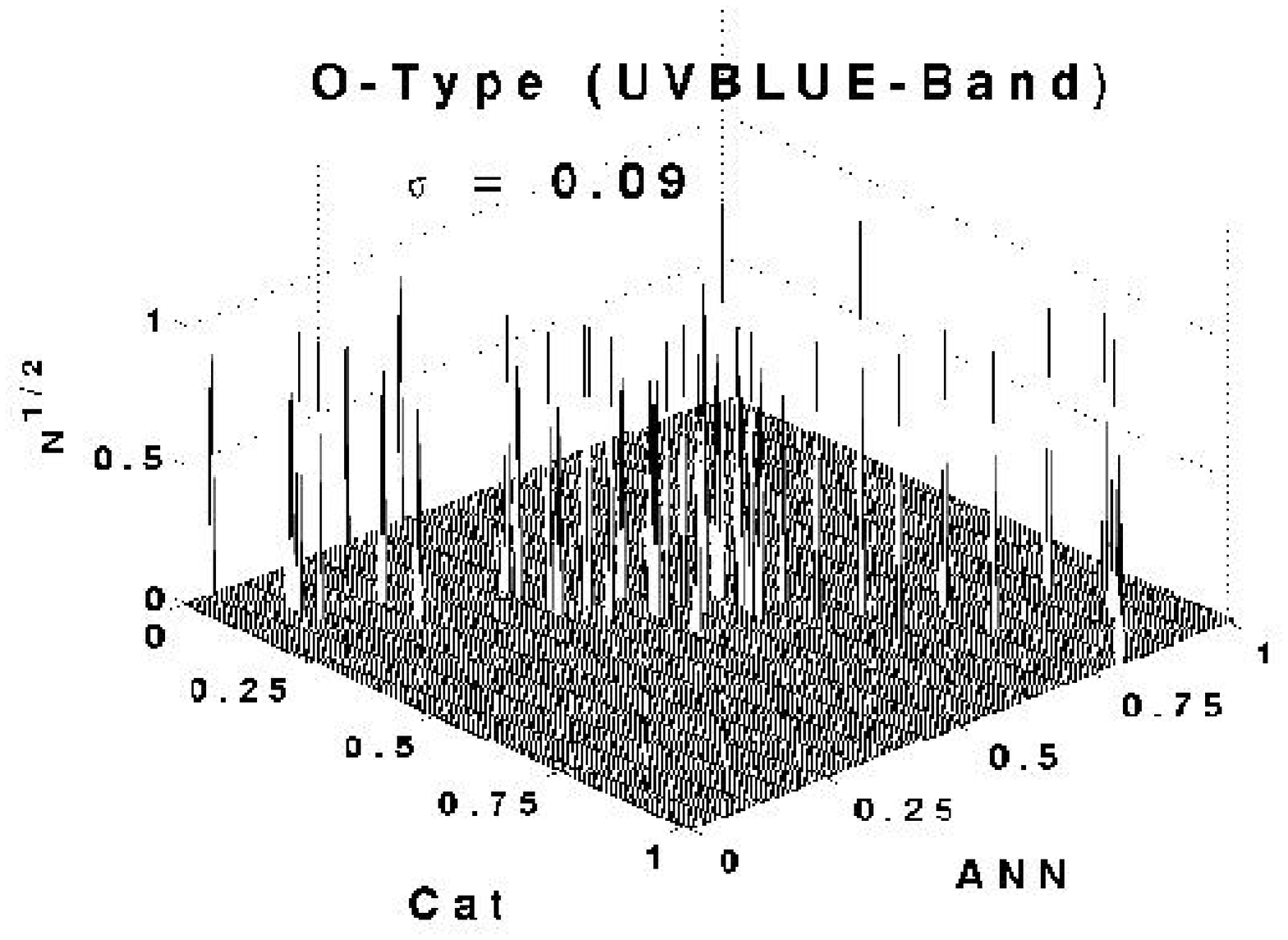}{}
\boxfig{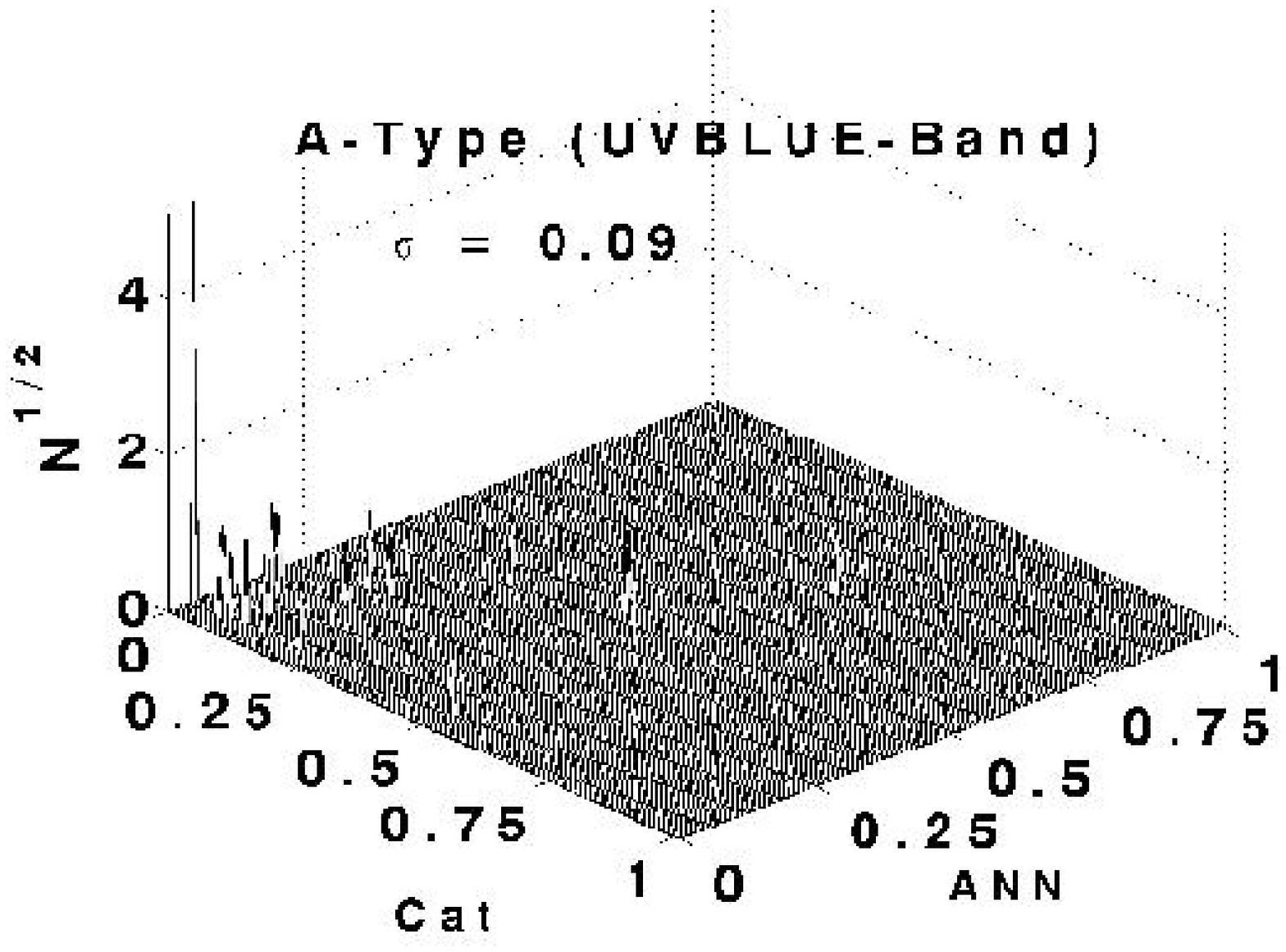}{}
\hspace{0.05in}
\boxfig{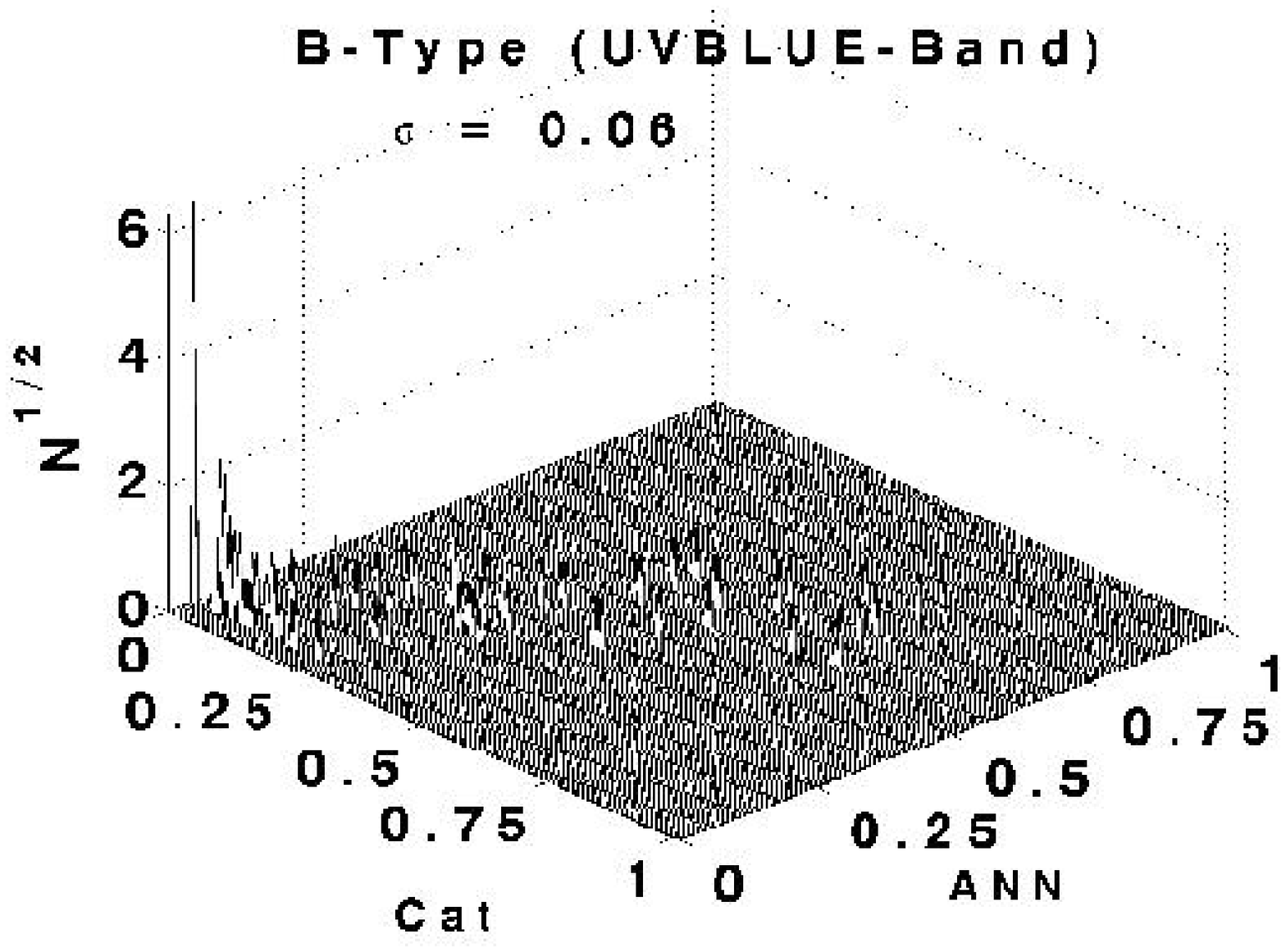}{}
\boxfig{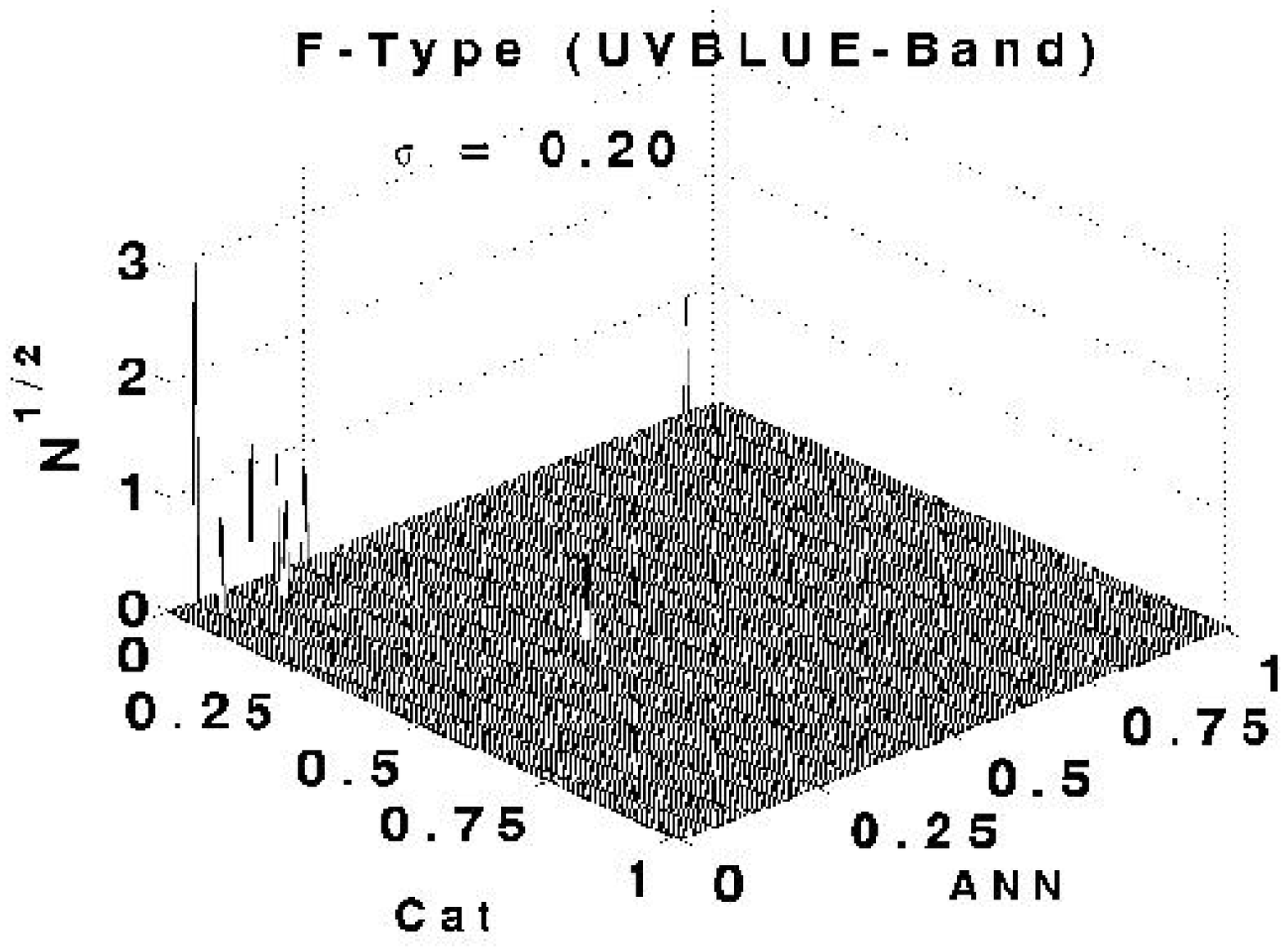}{}
\caption{Scatter plot of classification of 229 IUE stars (pre-classified
 into O, B, A and F spectral types) with UVBLUE bands  for colour excess estimates. The
 classification accuracy values $\sigma$ are shown for each case in units of
E(B-V) magnitudes.}
\end{figure}

\begin{figure}
\centering
\boxfig{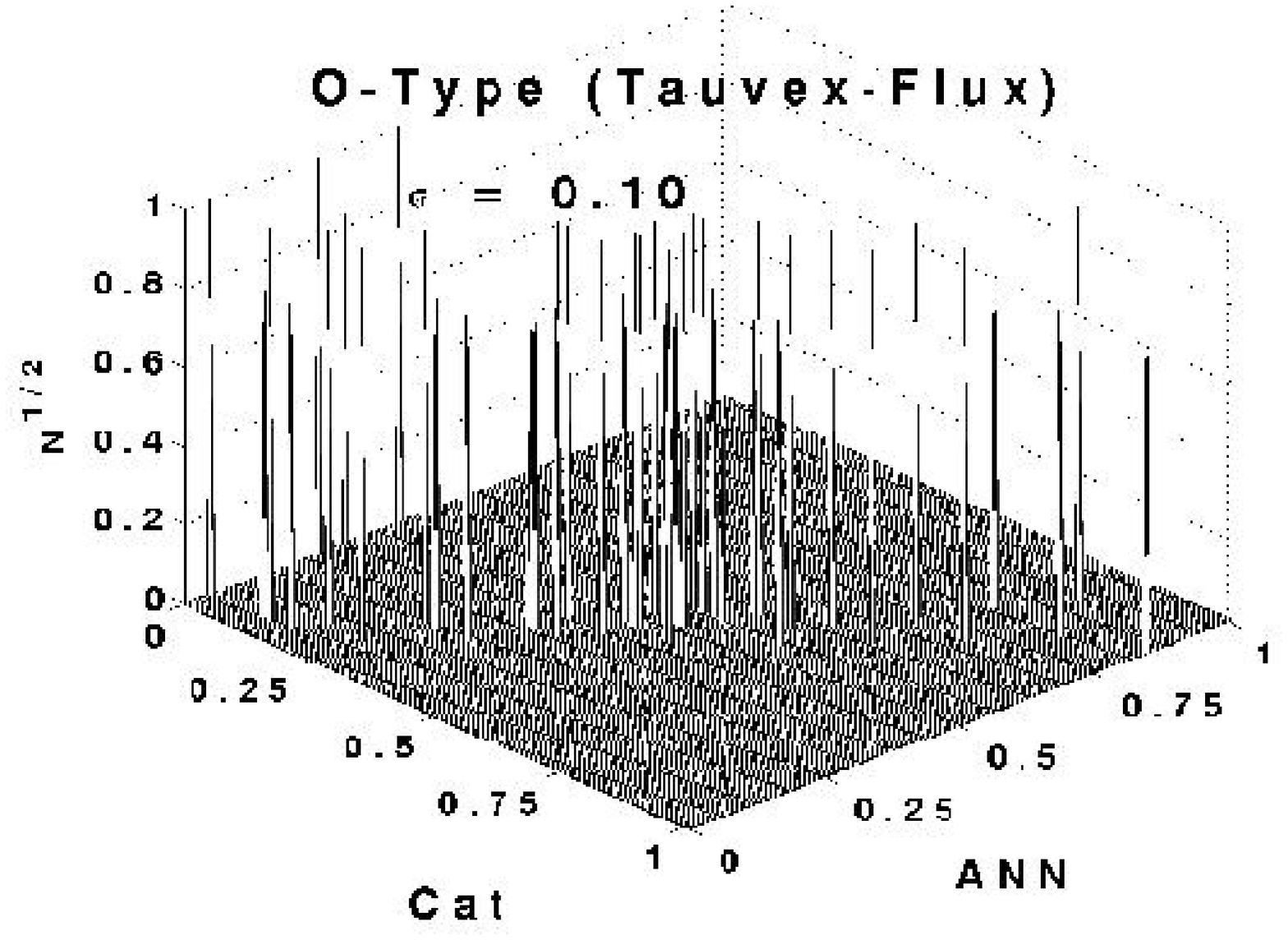}{}
\boxfig{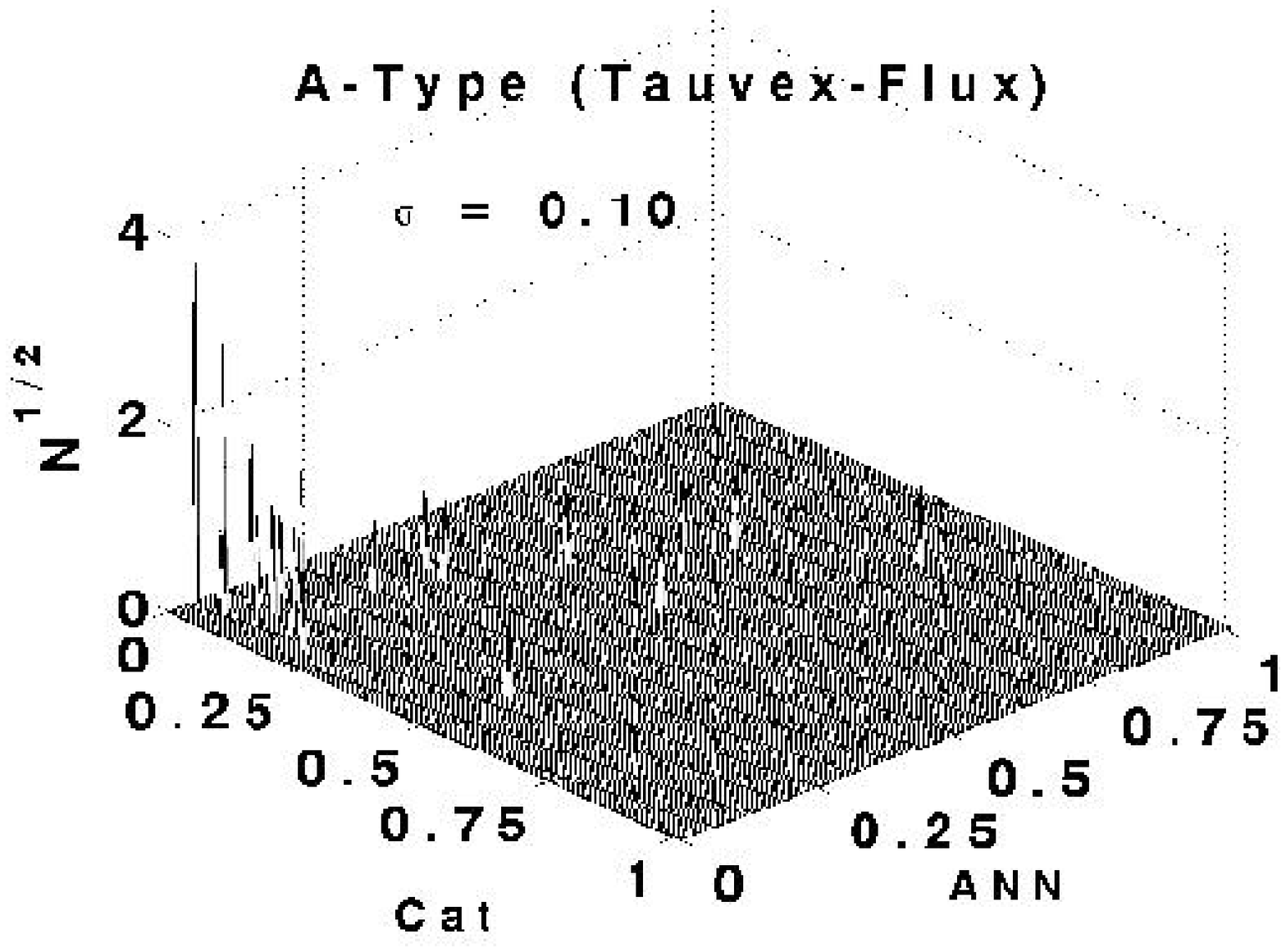}{}
\hspace{0.05in}
\boxfig{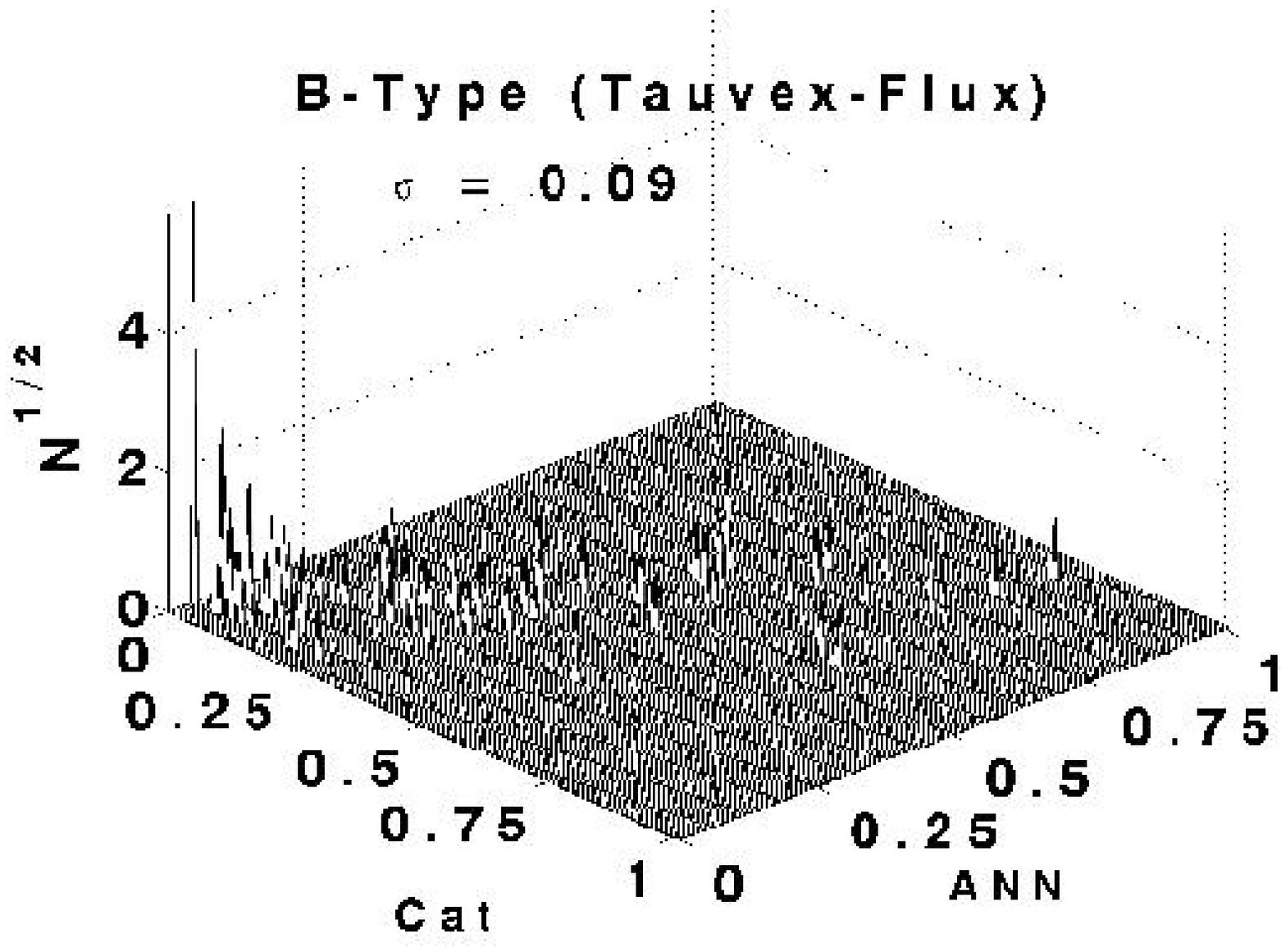}{}
\boxfig{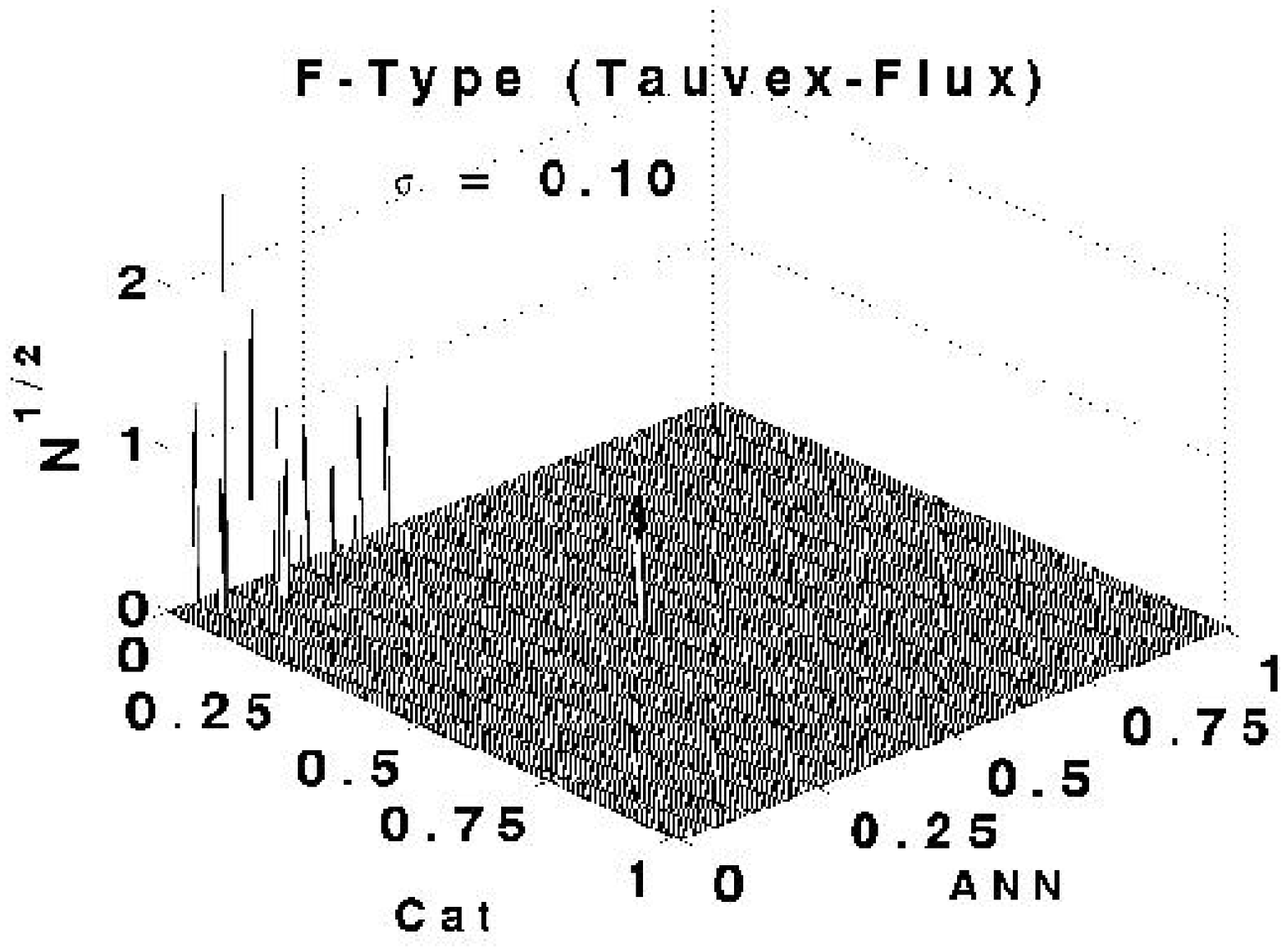}{}
\caption{Scatter plot of classification of 229 IUE stars (pre-classified
 into O, B, A and F spectral types) with TAUVEX fluxes for colour excess estimates. The
 classification accuracy values $\sigma$ are shown for each case in units of
E(B-V) magnitudes.}
\end{figure}

\begin{figure}
\centering
\boxfig{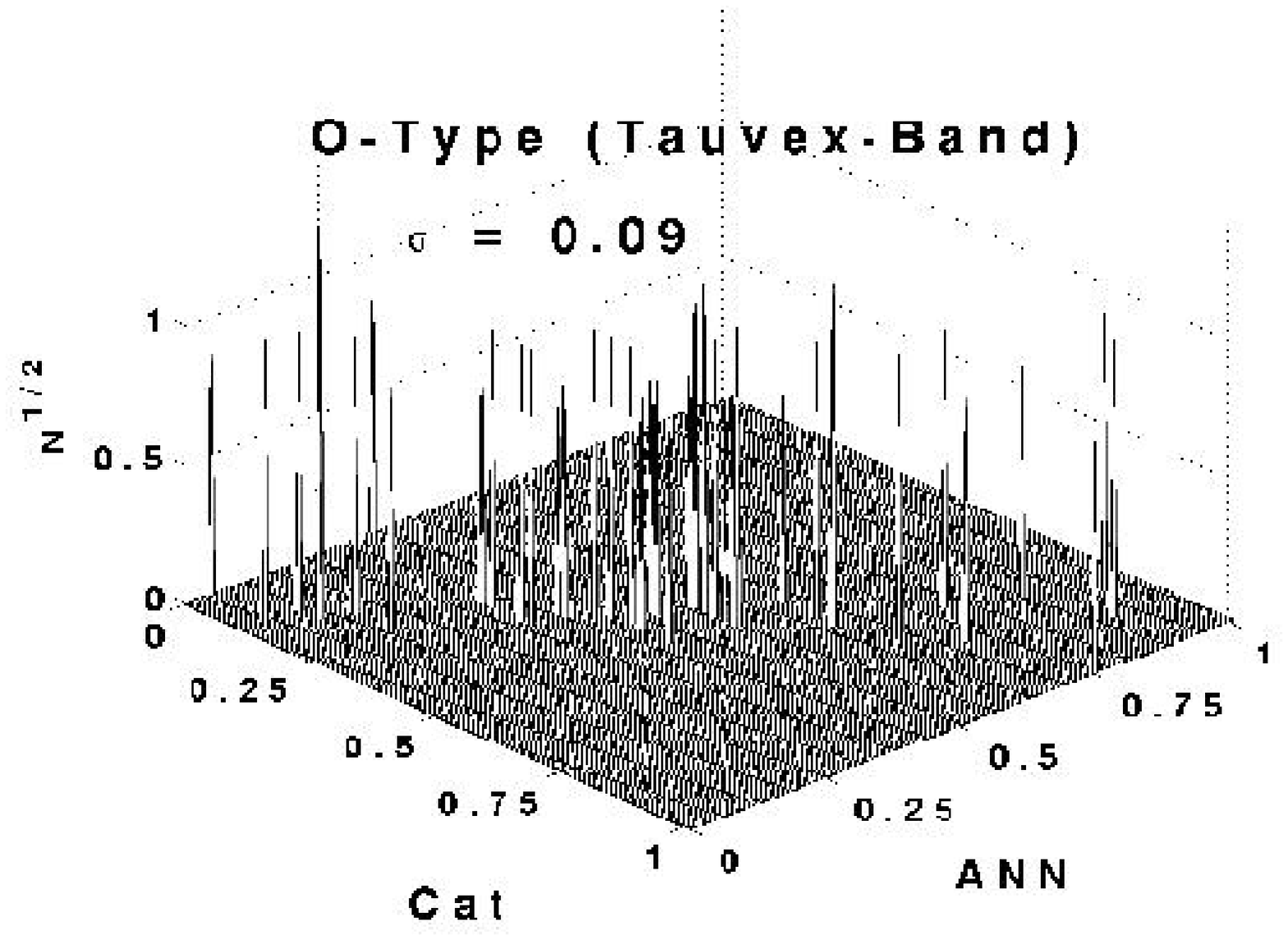}{}
\boxfig{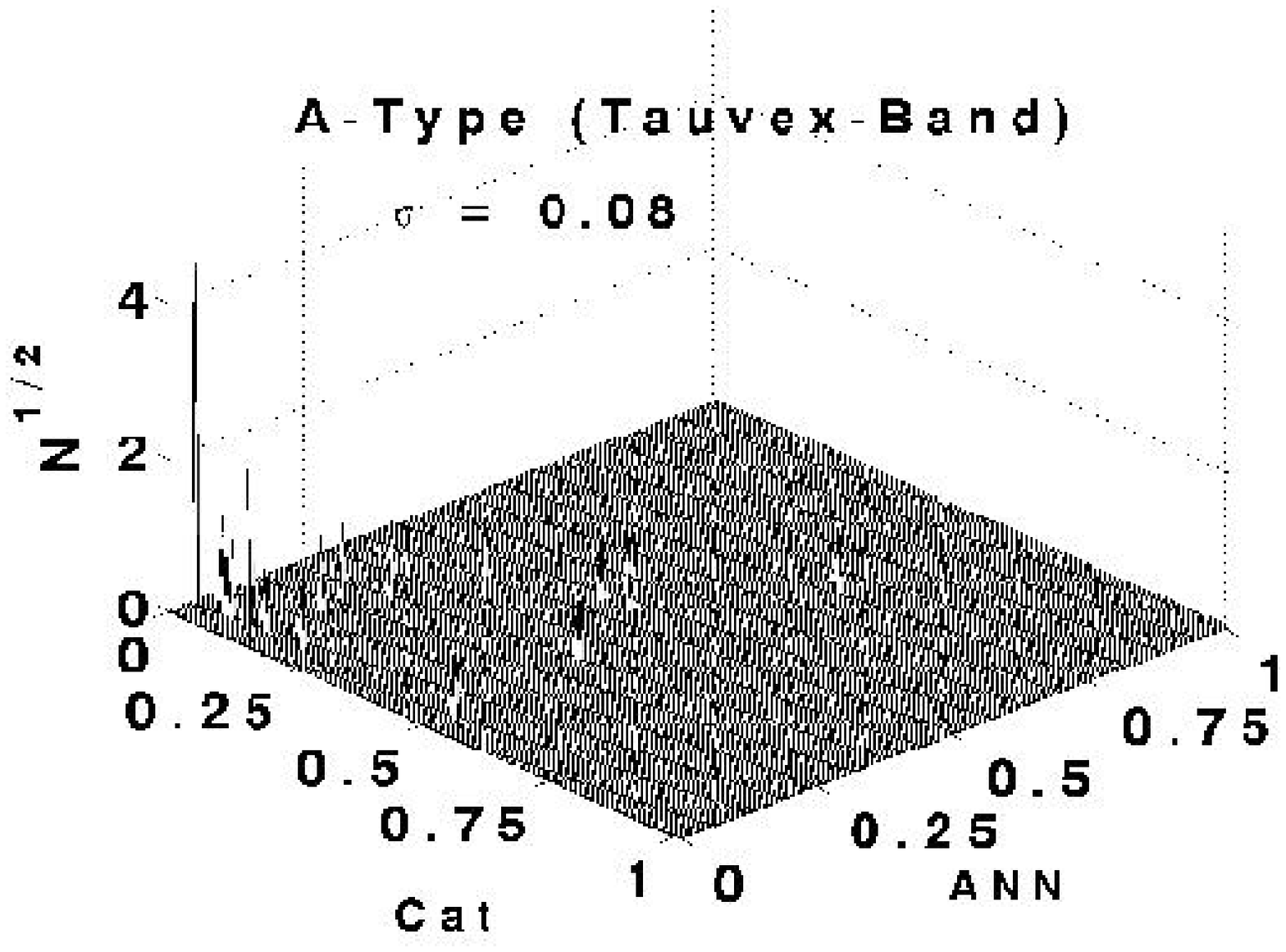}{}
\hspace{0.05in}
\boxfig{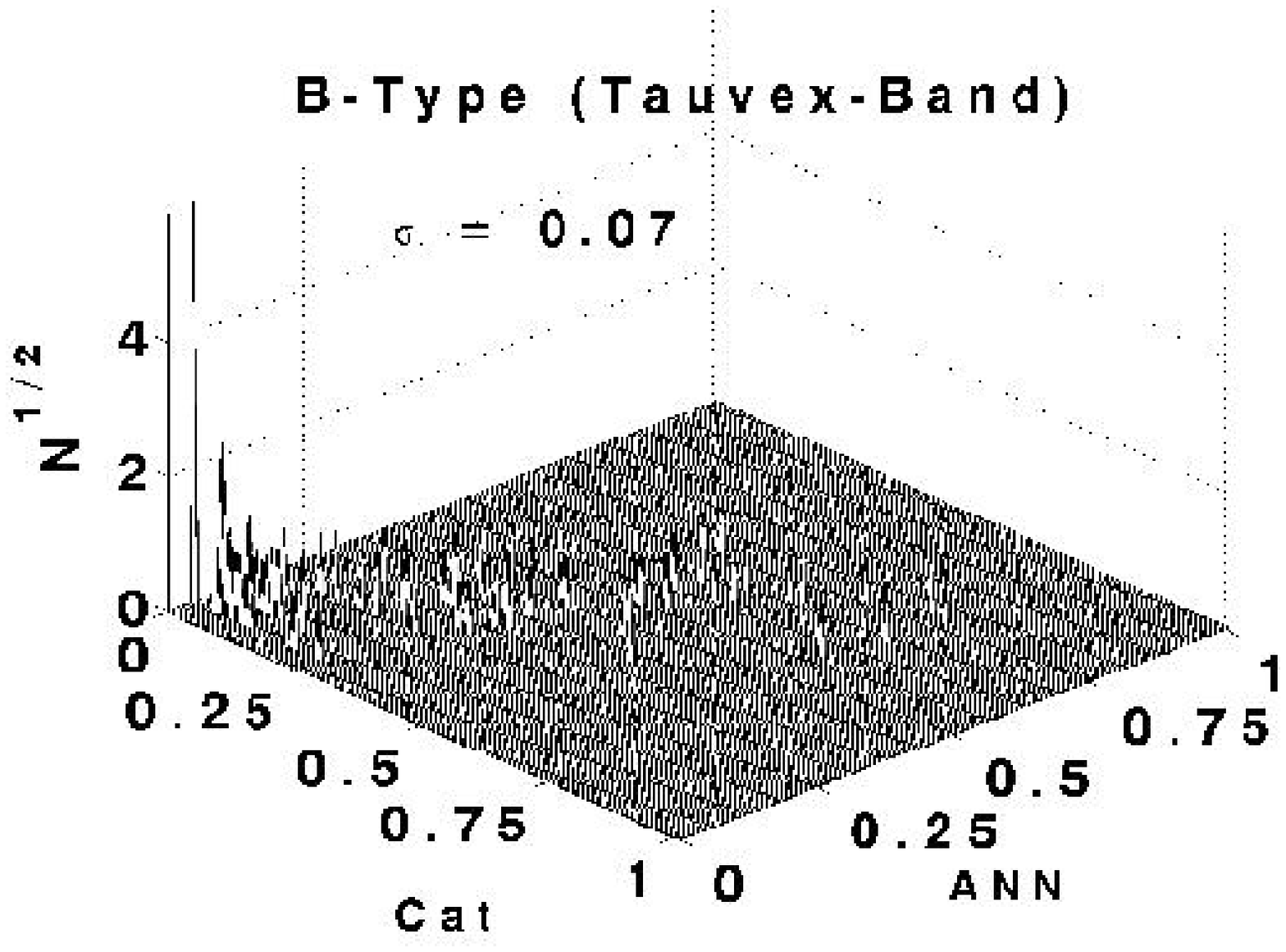}{}
\boxfig{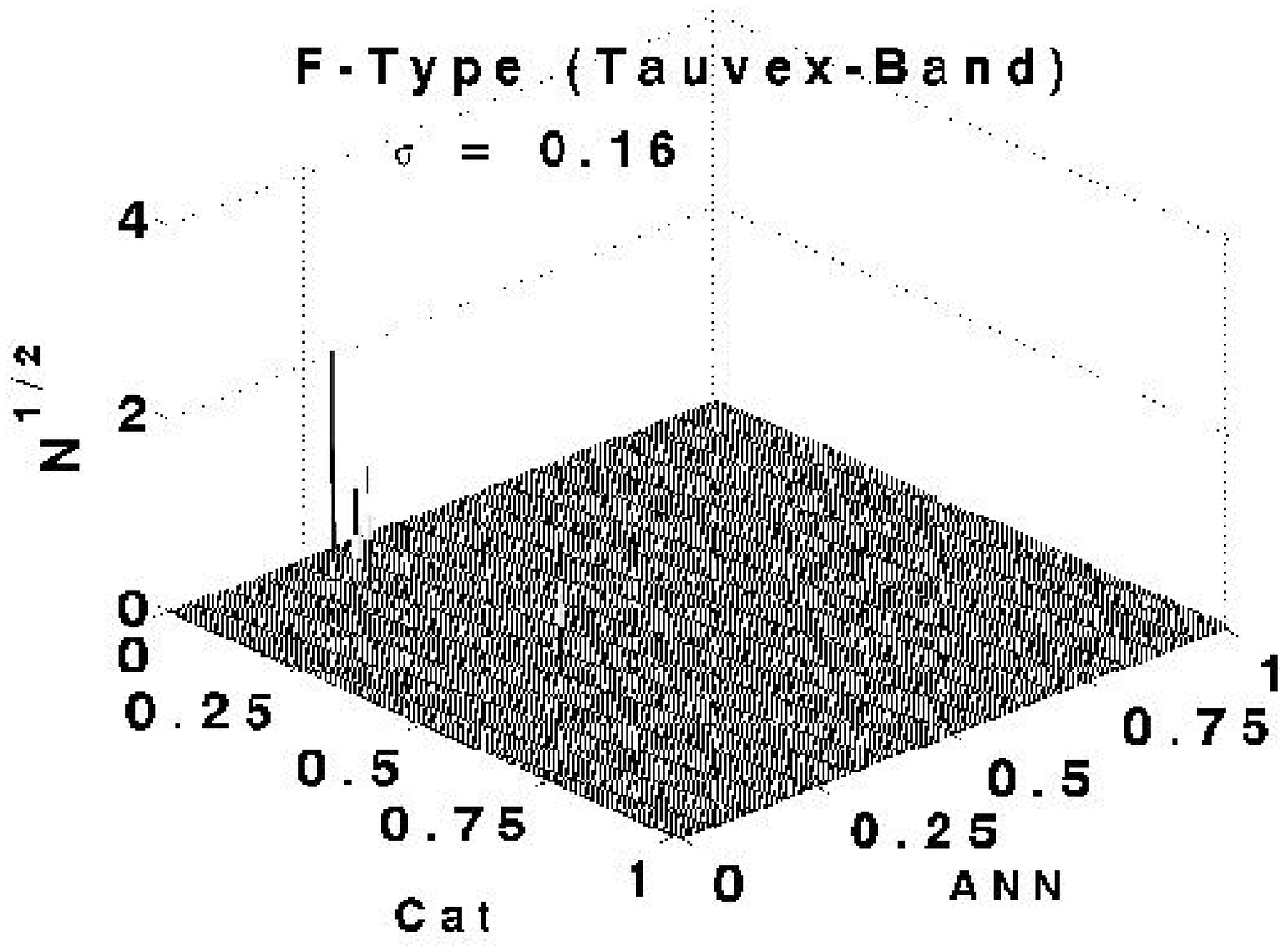}{}
\caption{Scatter plot of classification of 229 IUE stars (pre-classified
 into O, B, A and F spectral types) with TAUVEX fluxes for colour excess estimates. The
 classification accuracy values $\sigma$ are shown for each case in units of
E(B-V) magnitudes.}
\end{figure}

From the Figs. 9, 10, 11 \& 12 we see an overall colour excess estimate accuracy 
in the range of 0.20 in the worst case of F-Type spectra with bands to 0.06 in 
the best case for B-Type spectra with bands. 
The results with bands show better accuracies in comparison to the fluxes which 
may indicate that band data is a better estimator for colour excess than the fluxes.

The ANN inputs take most of the information in terms of 
absorption features which are embedded in the full range of
spectral fluxes (or the integrated fluxes in the band data)
for performing the classification. This information is
available for the hot stars like O, B and A but lacks in F or later spectral 
types. Due to this reason, the ANNs do not provide a good estimate of reddening
for these late type stars. Thus we have not estimated the colour excess for the 
G and K Type IUE spectra (the 3 nos of G Type and 1 no of K Type of the IUE 
test set mentioned in Table 1 have no reddening). 
Table 2 summarizes the results for both spectral type classification and colour 
excess estimation.

\begin{table*}
\begin{center}
\caption{Summary of Classification results.}
\begin{tabular}{lcccc}
\hline
Spectral Classification Error & & & &\\
$\sigma$(sub-spectral type) & & & &\\
\hline
Simulated Source: & TAUVEX  &   & UVBLUE & \\
 & Flux & Band & Flux & Band\\
\hline
& 3.97 & 3.84 & 3.77 & 3.39\\
\hline
Colour Excess E(B-V) Error & & & &\\
$\sigma$(magnitudes) & & & &\\
\hline
O-Type & 0.10 & 0.09 & 0.11 & 0.09\\
B-Type & 0.09 & 0.07 & 0.08 & 0.06\\
A-Type & 0.10 & 0.08 & 0.14 & 0.09\\
F-Type & 0.10 & 0.16 & 0.18 & 0.20\\
\hline
\end{tabular}
\end{center}
\end{table*}

\section[]{Conclusions}

Till now several studies have demonstrated that the artificial neural network schemes can reliably and successfully classify stellar spectral data as well as extract fundamental stellar
parameters in the visible region. The extension of applicability of this scheme to UV region has been less prevalent mainly because of non-availability of abundant data in this region. Nevertheless, some attempts have been made in the past to automate the process of classification of spectral data from the IUE satellite. In this paper, we have demonstrated that the artificial neural networks can be successfully employed to classify stellar photometric (band) data.

We have shown that the ANN tools developed by us can successfully classify the 
229 IUE spectra reduced to the four TAUVEX bands to an accuracy in the 
range of 3-4 sub-spectral types. We have also estimated the colour excess 
for the hot stars (O, B and A types) to an accuracy of  up to 0.1 magnitudes in 
terms of E(B-V) colours. Thus, even with the limitation of data from just 
photometric bands, ANNs have not only classified the stars, but also 
provided satisfactory estimates for interstellar extinction.

We hope that our automated pipeline will be used extensively to extract and 
validate data from virtual observatories as well as for the upcoming satellite 
data base expected from the TAUVEX and also the ASTROSAT and GAIA missions where 
one will be able to provide the interstellar extinction maps of our galaxy and 
which in turn could be modeled for dust distribution (Vaidya et al. 2001, 
Gupta et al. 2005, Vaidya et al. 2007).

\section*{Acknowledgments}

This work is supported by a Grant from Indian Space Research Organization under its RESPOND scheme. 
We would like to acknowledge an email correspondence with Prof. E. Bertone, 
for providing the relevant references which helped us in matching the UVBLUE parameter space to 
spectral types and luminosity classes.
AB thanks Dr. Kalpana Duorah for inspiring her to undertake this project.

\label{lastpage}
\end{document}